\documentclass[reqno]{amsart}
\usepackage{color}
\usepackage[mathscr]{eucal}
\usepackage{accents}
\usepackage[margin=2.5cm]{geometry}

\newtheorem{theorem}{Theorem}[section]
\newtheorem{lemma}[theorem]{Lemma}
\newtheorem{proposition}[theorem]{Proposition}
\newtheorem{corollary}[theorem]{Corollary}

\newtheorem{definition}{Definition}[section]
\newtheorem{remark}[theorem]{Remark}

\newcommand{\ut}{\widetilde u}
\newcommand{\uh}{\widehat u}

\newcommand{\uth}{\widehat{\widetilde u}}
\newcommand{\uo}{\overline u}

\newcommand{\cUt}{\widetilde\cU}
\newcommand{\cUh}{\widehat\cU}

\newcommand{\cUo}{\overline{\cU}}

\newcommand{\wt}{\widetilde}
\newcommand{\wh}{\widehat}
\newcommand{\ol}{\overline}
\newcommand{\mr}{\mathring}

\newcommand{\vp}{\varphi}

\newcommand{\rvp}{\mathring{\varphi}}
\newcommand{\rmpsi}{\mathring{\psi}}
\newcommand{\rmp}{\mathring{p}}
\newcommand{\rmq}{\mathring{q}}
\newcommand{\up}{\Upsilon}
\newcommand{\lam}{\Lambda}
\newcommand{\az}{\mathtt A}
\newcommand{\bz}{\mathtt B}
\newcommand{\pot}{\upsilon}

\newcommand{\bphi}{\boldsymbol \phi}

\newcommand{\rlam}{\mathring{\Lambda}}
\newcommand{\rup}{\mathring{\Upsilon}}

\newcommand{\cA}{\mathscr A}
\newcommand{\cB}{\mathscr B}
\newcommand{\cC}{\mathscr C}
\newcommand{\cD}{\mathscr D}
\newcommand{\cF}{\mathcal F}
\newcommand{\cR}{\mathcal R}
\newcommand{\cK}{\mathscr K}

\newcommand{\cS}{\mathscr S}
\newcommand{\cU}{\mathscr U}

\newcommand{\cM}{\mathscr M}
\newcommand{\cH}{\mathscr H}

\newcommand{\cP}{\mathscr P}
\newcommand{\cZ}{\mathscr Z}
\newcommand{\cT}{\mathscr T}
\newcommand{\pp}{\mathfrak p}
\newcommand{\qq}{\mathfrak q}

\newcommand{\asf}{\accentset{\boldsymbol \frown}}
\newcommand{\fru}{\asf{u}}
\newcommand{\frU}{\asf{\cU}}

\DeclareMathAccent{\wtilde}{\mathord}{largesymbols}{"65}
\DeclareMathAccent{\what}{\mathord}{largesymbols}{"62}

\numberwithin{equation}{section}
\begin{document}
\title{Multidimensional Inverse Scattering of Integrable Lattice Equations}
\author{Samuel Butler}
\address{School of Mathematics and Statistics F07, The University of Sydney, NSW 2009 Australia}
\email{s.butler@sydney.edu.au}
\date{23 February 2012}

\begin{abstract}

We present a discrete inverse scattering transform for all ABS equations excluding Q4. The nonlinear partial difference equations presented in the ABS hierarchy represent a comprehensive class of scalar affine-linear lattice equations which possess the multidimensional consistency property. Due to this property it is natural to consider these equations living in an N-dimensional lattice, where the solutions depend on N distinct independent variables and associated parameters.  The direct scattering procedure, which is one-dimensional, is carried out along a staircase within this multidimensional lattice. The solutions obtained are dependent on all N lattice variables and parameters. We further show that the soliton solutions derived from the Cauchy matrix approach are exactly the solutions obtained from reflectionless potentials, and we give a short discussion on inverse scattering solutions of some previously known lattice equations, such as the lattice KdV equation.

\end{abstract}
\maketitle
\section{Introduction}\label{sec:intro}

The Inverse Scattering Transform (IST) has been widely used as a mathematical tool for obtaining solutions of integrable nonlinear partial differential equations of two independent variables, such as the Korteweg-de Vries equation (KdV), the sine-Gordon equation and the nonlinear Schr\"odinger equation (see e.g. Ablowitz and Segur \cite{as:81}). One uses the IST to solve the initial-value problem, where an integrable initial condition is usually given on the line $t=0$, the Jost solutions are constructed, the time-dependent scattering data is computed, and the full time-dependent solution is then reconstructed. The IST naturally describes soliton solutions, where for the case of the KdV these correspond to reflectionless potentials. A rigorous discussion of the direct scattering of the time-independent Schr\"odinger equation along the full line can be seen in Deift and Trubowitz \cite{dt:79} and the case of the half-line in Fokas et al. \cite{fis:05}. 

The discrete IST dates back to the 1970's (to the best of the author's knowledge) with the works of Case and Kac \cite{ck:73} \cite{c:73}, where the authors solved a direct discretisation of the Schr\"odinger equation on the half-line $n\geq0$. Subsequently Flaschka \cite{f:74} looked at the addition of continuous time evolution to the discrete IST in order to solve differential-difference equations arising from the Toda lattice. Other applications of the discrete IST to differential-difference equations can be seen in \cite{pl:82}, \cite{b:99} and \cite{abp:07}. A new scattering problem was derived in \cite{s:99} from an ``exact" discretisation of the Schr\"odinger equation. This problem has since been studied by various authors \cite{bpps:01} \cite{r:02} \cite{lp:07} for the cases of continuous and discrete time evolution, and as an eigenvalue problem for analytic difference operators. In 2010 a rigorous formulation of the discrete IST for this equation was given in \cite{bj:10}, where the forward scattering problem was carried out on the line $n=0$ and new sufficient conditions were derived for the required properties of the transmission and reflection coefficients.

In the present paper we build on the foundations of the machinery in \cite{bj:10} to present a discrete IST for all ABS equations \cite{abs:03}, excluding Q4, in a complex multidimensional setting. The ABS list consists of nine partial difference equations with multidimensional consistency \cite{nw:01}, \cite{n:02}, each of which representing a class of partial difference equations through point transformations of parameters and M\"obius transformations of the dependent variable. The list consists of H-type, Q-type and A-type equations. The H-type equations
\begin{subequations}\label{eq:Hlist}
\begin{align}
&{\rm H}1: \;\;\; (u-\uth)(\ut-\uh)+p^2-q^2=0 \\
&{\rm H}2: \;\;\; (u-\uth)(\ut-\uh)+(p^2-q^2)(u+\ut+\uh+\uth)-p^4+q^4=0 \\
&{\rm H}3_\delta: \;\;\; Q(u\ut+\uh\uth)-P(u\uh+\ut\uth)+\delta\left(\frac{p^2-q^2}{PQ}\right)=0 \\
&\hspace{0.5in} {\rm where} \; P^2=a^2-p^2, \; Q^2=a^2-q^2 \nonumber
\end{align}
\end{subequations}
are all covered in this paper's scope. Here we have adopted the standard notation
\[
u:=u_{n,m}, \;\;\;\;\; \ut:u_{n+1,m}, \;\;\;\;\; \uh=u_{n,m+1}, \;\;\;\;\; \uth=u_{n+1,m+1}
\]
where $n$ and $m$ ($\in\mathbb{Z}$) are the independent variables of a two-dimensional lattice, with associated parameters $p$ and $q$ respectively. Of the four equations of Q-type we consider all except the topmost Q4, that is
\begin{subequations}\label{eq:Qlist}
\begin{align}
&{\rm Q}1_\delta: \;\;\; \rmp(u-\uh)(\ut-\uth)-\rmq(u-\ut)(\uh-\uth)+\delta^2\rmp\rmq(\rmp-\rmq)=0 \\
&{\rm Q}2: \;\;\;\rmp(u-\uh)(\ut-\uth)-\rmq(u-\ut)(\uh-\uth) \\
&\hspace{0.3in} +\rmp\rmq(\rmp-\rmq)(u+\ut+\uh+\uth)-\rmp\rmq(\rmp-\rmq)(\rmp^2-\rmp\rmq+\rmq^2)=0 \\
\label{eq:Q3}
&{\rm Q}3_\delta: \;\;\; P(u\uh+\ut\uth)-Q(u\ut+\uh\uth)-(p^2-q^2)\left((\ut\uh+u\uth)+\frac{\delta^2}{4PQ}\right)=0,
\end{align}
\end{subequations}
where $\rmp=\frac{a^2}{p^2-a^2}$, $\rmq=\frac{a^2}{q^2-a^2}$ and the parameters $\pp:=(p,P)$, $\qq:=(q,Q)$ lie on the Jacobi elliptic curve $\{(x,X):X^2=(x^2-a^2)(x^2-b^2)\}$. The two remaining A-type equations are related by straightforward gauge transformations to ${\rm Q}1_\delta$ and ${\rm Q}3_o$, and as such are not mentioned here as distinct examples. The choice of parametrisation that we have used was presented in \cite{nah:09}, in which each equation is parametrised by the {\it common} lattice parameters $p$ and $q$, which occur naturally in all their solutions. This is advantageous for the degeneration procedure carried out in Section \ref{sec:degen}. 

Rather than simply considering the equations \eqref{eq:Hlist} and \eqref{eq:Qlist} as equations imposed on a two-dimensional lattice, the  multidimensional consistency suggests that the most natural way to consider these equations is in an N-dimensional lattice with independent variables $n_1$, ..., $n_{\rm N}$ and parameters $p_1,$, ..., $p_{\rm N}$. That is to say we require that each equation hold on every elementary quadrilateral throughout all N lattice directions, and thus the solution $u$ of each equation will have dependence on each independent variable $n_1$, ..., $n_{\rm N}$ and each parameter $p_1$, ..., $p_{\rm N}$. This is the space in which we carry out the inverse scattering transform, and in Section \ref{sec:ist} we rigorously develop the machinery for this, for the case of ${\rm Q}3_\delta$. Here the initial condition is taken on an infinite one-dimensional staircase within this N-dimensional space. In Section \ref{sec:degen} we perform the degeneration procedure between the Q-type and H-type equations, resulting in multidimensional solutions for all equations in the lists \eqref{eq:Hlist} and \eqref{eq:Qlist}. In Section \ref{sec:arb} we consider the case of an arbitrary reflectionless potential, and in Section \ref{sec:others} we give explicit solutions to some previously known lattice equations, including the lattice KdV equation, which is not multidimensionally consistent. Finally we mention that in \cite{av:04} it was explained how a well-posed initial-value problem for an arbitrary finite quad-graph can be immersed into a canonical initial-value problem for an N-dimensional unit cube. Thus the initial-value problem that we solve also solves that for an arbitrary two-dimensional quad-graph.

\section{Inverse Scattering Transform for ${\rm Q}3_\delta$}\label{sec:ist}

In this section we develop the scattering transform for ${\rm Q}3_\delta$, where we assume that the solution $u$ and all parameters are complex. We consider the equation being imposed on each quadrilateral of an N-dimensional lattice, with independent variables $n_1$, ..., $n_N$ and associated parameters $\pp_1$, ..., $\pp_N$. For example if ${\rm N}=3$ with independent variables $n,m,l$ and parameters $p,q,r$ respectively, then we impose that \eqref{eq:Q3} hold in the $(n,m)$-plane with parameters $p$ and $q$, in the $(n,l)$-plane with parameters $p$ and $r$ and in the $(m,l)$-plane with parameters $q$ and $r$. This is permissible due to the multidimensional consistency of the equation, which is a well-known feature of partial difference equations, and has no analogue in the continuous case. 

\subsection{Linear Problem for ${\rm Q}3_\delta$} $\\$

Consider a particular direction within the N-dimensional lattice, with independent variable $n_k$ and associated parameter $\pp_k$. Along this direction we then define the biquadratic \cite{abs:03} $\cH$ associated with \eqref{eq:Q3} to be
\begin{equation}\label{eq:Hp}
\cH_{\pp_k}(u,\fru):=P_k(u^2+\fru^2)-(2p_k^2-a^2-b^2)u\fru+\frac{\delta^2}{4P_k},
\end{equation}
where $\fru$ denotes a shift of $u$ in the $n_k$-direction. This biquadratic plays a central role in the linearisation of \eqref{eq:Q3}, and in fact a convenient means of expressing \eqref{eq:Hp} is by defining the function $\cU$ through the first-order equation
\begin{equation}\label{eq:Hfac}
\cU\frU=\cH_{\pp_k}(u,\fru)=P_k(u^2+\fru^2)-(2p_k^2-a^2-b^2)u\fru+\frac{\delta^2}{4P_k}.
\end{equation}
Here $\cU$ is determined up to a multiplicative factor of $\alpha^{(-1)^{n_k}}$ for some arbitrary constant $\alpha$, and by imposing $\alpha=1$, $\cU$ is determined uniquely. A Lax equation (first given in \cite{na:10}) in the $n_k$ direction for ${\rm Q}3_\delta$ may then be written as
\begin{equation}\label{eq:Nlax}
(p_k^2-\zeta^2)^{\frac12}\asf{\bphi}=\frac1{\cU}\left(\begin{array}{cc} P_k\fru-(p_k^2-b^2)u & \zeta^2-b^2 \vspace{0.1 in}\\ \cU\frU-\frac{\delta^2(p_k^2-b^2)}{4P_k(\zeta^2-b^2)} & (p_k^2-b^2)\fru-P_ku \end{array}\right)\bphi.
\end{equation}
Here $\zeta$ is the spectral parameter. Now since we could have chosen any one of the possible N directions we in fact have N biquadratics of the form \eqref{eq:Hp} and N Lax equations of the form \eqref{eq:Nlax}. The dependence of $\cU$ on each of the N independent variables is defined through \eqref{eq:Hfac}. This in turn gives rise to $\frac12$N(N-1) Lax pairs, which are consistent (in the sense $\wh{\wt\bphi}=\wt{\wh\bphi}$) if and only if $u$ satisfies \eqref{eq:Q3} in each pair of lattice directions. 

\subsection{Initial Staircase}\label{subsec:IS} $\\$

We now consider the initial-value space for the direct scattering problem. Let $\Gamma$ denote an infinite multidimensional staircase within the N-dimensional lattice, which is always non-decreasing in every lattice direction, and is defined through some stepping algorithm. An example for N=3 is a (1,2,4) staircase, that is one step in the $n$-direction followed by two steps in the $m$-direction followed by four steps in the $l$-direction, and then repeating this process indefinitely. The staircase will then exist in a subspace (and span every direction of this subspace) of the N-dimensional lattice, and we denote this collection of lattice directions by I.
\par
Given the staircase $\Gamma$ described above, we may enumerate the points along this staircase by a new independent variable $i$, and the parameter associated with each shift along $\Gamma$ by $\pp_i=(p_i,P_i)$. Thus each iteration $i\to i+1$ will in fact correspond to an iteration in one of the N lattice variables, for example $n_3\to n_3+1$ with parameter $\pp_3$. We also use the notation $\ol\bphi$ to denote a shift of $\bphi$ along the staircase.
\par
By \eqref{eq:Nlax} we therefore have, along $\Gamma$, a sequence of Lax equations
\begin{equation}\label{eq:ds''}
(p_i^2-\zeta^2)^{\frac12}\ol{\bphi}=\frac1{\cU}\left(\begin{array}{cc} P_i\uo-(p_i^2-b^2)u & \zeta^2-b^2 \vspace{0.1 in}\\ \cU\cUo-\frac{\delta^2(p_i^2-b^2)}{4P_i(\zeta^2-b^2)} & (p_i^2-b^2)\uo-P_iu \end{array}\right)\bphi.
\end{equation}
(Note that the parameters $a$ and $b$, which appear originally in the elliptic parametrisation of ${\rm Q}3_\delta$, are constant for all lattice directions, and as such do not change along $\Gamma$.) This sequence of linear equations along $\Gamma$ defines the direct scattering problem. Even though this initial-value space exists in up to N dimensions, the staircase itself, and thus the nature of the direct scattering problem, is {\it one-dimensional}. Allowing $\Gamma$ to exist in multiple dimensions changes the parameter values along the staircase, but does not change the dimension of the direct scattering problem. As an analogy to the IST for continuous equations, the remaining N-1 lattice variables play the role of N-1 discrete ``time" variables.
\par
Rather than consider the matrix problem \eqref{eq:ds''} it is convenient to consider the second-order equation satisfied by the first component of $\bphi$, which one can calculate to be
\begin{equation}\label{eq:phi0}
(p_{i+1}^2-\zeta^2)^{\frac12}\ol{\ol\phi}_1-\left(\frac{P_{i+1}\ol\uo-(p_{i+1}^2-p_i^2)\uo-P_iu}{\cUo}\right)\ol{\phi}_1+(p_i^2-\zeta^2)^{\frac12}\phi_1=0.
\end{equation}
The direct scattering problem for ${\rm Q}3_\delta$ is the solving of equation \eqref{eq:phi0}, which will be discussed in due course.

\subsection{Boundary Conditions for $u$ and $\cU$} $\\$

Soliton solutions to $Q3_\delta$ depending on an arbitrary number of independent variables and parameters are given in \cite{nah:09}. These are closely related to solutions of the NQC equation \cite{nqc:83}. These solutions suggest that the most natural ansatz for the solution of $Q3_\delta$ is
\begin{equation}\label{eq:uform}
u=\cA\cF(a,b)\cS_\cA+\cB\cF(a,-b)\cS_\cB+\cC\cF(-a,b)\cS_\cC+\cD\cF(-a,-b)\cS_\cD.
\end{equation}
where the exponential functions $\cF$ are given by
\begin{equation}\label{eq:F}
\cF(a,b)=\prod_{r=1}^N\left(\frac{(p_r+a)(p_r+b)}{(p_r-a)(p_r-b)}\right)^{\frac12n_r},
\end{equation}
and the four constants $\cA$, $\cB$, $\cC$ and $\cD$ satisfy the single constraint
\begin{equation}\label{eq:abcd}
\cA\cD(a+b)^2-\cB\cC(a-b)^2=-\frac{\delta^2}{16ab}.
\end{equation}
This final condition is necessary and sufficient for the background solution obtained by setting $\cS_\cA=\cS_\cB=\cS_\cC=\cS_\cD=1$ to solve ${\rm Q}3_\delta$. Thus from the ansatz \eqref{eq:uform} the problem of determining the unknown quantity $u$ has been turned into that of determining the four unknown quantities $\cS_\cA$ etc. We now impose some boundary conditions on these functions, namely that they satisfy
\begin{align*}
\cS_\cA\sim1, \;\;\; \cS_\cB\sim1, \;\;\; \cS_\cC\sim1, \;\;\; \cS_\cD\sim1 \;\;\; {\rm as} \; i\to-\infty, \\
\cS_\cA\sim\cS_\cA^{o}, \;\;\; \cS_\cB\sim\cS_\cB^{o}, \;\;\; \cS_\cC\sim\cS_\cC^{o}, \;\;\; \cS_\cD\sim\cS_\cD^{o} \;\;\; {\rm as} \; i\to+\infty
\end{align*}
where $\cS_\cA^o, \cS_\cB^o, \cS_\cC^o$ and $\cS_\cD^o$ are constants satisfying $\cS_\cA^{o}\cS_\cD^{o}=\cS_\cB^{o}\cS_\cC^{o}=1$ (which ensures the background solution obtained by setting $\cS_\cA=\cS_\cA^{o}$, etc. does indeed solve ${\rm Q}3_\delta$.). This is the case for the soliton solutions \cite{nah:09} in which lower-order terms vanish exponentially, provided that we assume that the plane-wave factors appearing in the functions $\cS_\cA$, etc. do not have modulus 1.
\par
With the above boundary conditions imposed on the functions $\cS_\cA$ etc., and therefore on $u$, we now calculate the boundary behaviour of the coefficient of $\ol\phi_1$ in the direct scattering problem \eqref{eq:phi0}. From the boundary behaviour of $u$, namely
\begin{subequations}\label{eq:BCu}
\begin{align}
&u\sim\cA\cF(a,b)+\cB\cF(a,-b)+\cC\cF(-a,b)+\cD\cF(-a,-b) \; {\rm as} \; i\to-\infty, \\
&u\sim\cA\cF(a,b)\cS_\cA^{o}+\cB\cF(a,-b)\cS_\cB^{o}+\cC\cF(-a,b)\cS_\cC^{o}+\cD\cF(-a,-b)\cS_\cD^{o} \; {\rm as} \; i\to+\infty,
\end{align}
\end{subequations}
the boundary behaviour of $\cU$ is then determined by \eqref{eq:Hfac}. As $i\to-\infty$ one may calculate this to be
\begin{subequations}\label{eq:BCU}
\begin{equation}
\cU\sim(a+b)\cA\cF(a,b)+(a-b)\cB\cF(a,-b)-(a-b)\cC\cF(-a,b)-(a+b)\cD\cF(-a,-b),
\end{equation}
while as $i\to+\infty$,
\begin{equation}
\cU\sim(a+b)\cA\cF(a,b)\cS_\cA^{o}+(a-b)\cB\cF(a,-b)\cS_\cB^{o}-(a-b)\cC\cF(-a,b)\cS_\cC^{o}-(a+b)\cD\cF(-a,-b)\cS_\cD^{o}.
\end{equation}
\end{subequations}
Therefore by combining \eqref{eq:BCu} and \eqref{eq:BCU} one finds that at both ends of the staircase the numerator of the coefficient of $\ol\phi_1$ in \eqref{eq:phi0} factorises and we have
\begin{equation}\label{eq:bcpot}
\left(\frac{P_{i+1}\ol\uo-(p_{i+1}^2-p_i^2)\uo-P_iu}{\cUo}\right)\sim p_i+p_{i+1} \; {\rm as} \; i\to\pm\infty.
\end{equation}

\subsection{Forward Scattering Problem} $\\$

The forward scattering problem for ${\rm Q}3_\delta$ is the solving of equation \eqref{eq:phi0} for the function $\phi_1$. Due to the imposed boundary conditions on $u$ \eqref{eq:BCu}, which give rise to \eqref{eq:BCU} and \eqref{eq:bcpot}, we make the following definition.

\begin{definition}\label{def:pot}
The function $\pot$, which will be henceforth referred to as the {\bf potential}, is defined to be
\begin{equation}\label{eq:upot}
\pot_{i+1}\equiv\ol\pot:=\left(\frac{P_{i+1}\ol\uo-(p_{i+1}^2-p_i^2)\uo-P_iu}{\cUo}\right)-p_i-p_{i+1}.
\end{equation}
\end{definition}

From \eqref{eq:bcpot} we clearly have $\pot\to0$ as $i\to\pm\infty$. Equation \eqref{eq:phi0}, which characterises the forward scattering problem for ${\rm Q}3_\delta$, then becomes
\begin{equation}\label{eq:phi1}
(p_{i+1}^2-\zeta^2)^{\frac12}\phi_1(i+2;\zeta)-(p_i+p_{i+1}+\pot_{i+1})\phi_1(i+1;\zeta)+(p_i^2-\zeta^2)^{\frac12}\phi_1(i;\zeta)=0.
\end{equation}
The potential $\pot_i$ is in fact the initial profile that must be specified before carrying out the forward scattering problem. As will be seen below this must satisfy a particular summability condition. Even though by \eqref{eq:upot} this is equivalent to giving some nonlinear derivative of the actual solution $u$ to ${\rm Q}3_\delta$, we shall see that, with the appropriate boundary conditions, it is sufficient initial data to be able to reconstruct the solution $u$ through the inverse problem.

\subsection{Jost Solutions} $\\$

We now look at particular solutions to equation \eqref{eq:phi1}. 
\begin{definition}
The Jost solutions $\vp$, $\rvp$, $\psi$ and $\rmpsi$, which solve equation \eqref{eq:phi1}, are defined by the boundary conditions
\begin{equation}\label{eq:BCjost}
\left.\begin{array}{c} \vp(i;\zeta)\sim\prod_{r=i_o}^{i-1}\left(\frac{p_r+\zeta}{p_r-\zeta}\right)^{\frac12} \\ \rvp(i;\zeta)\sim\prod_{r=i_o}^{i-1}\left(\frac{p_r-\zeta}{p_r+\zeta}\right)^{\frac12} \end{array} \right\} \; {\rm as} \; i\to-\infty, \;\;\; \left. \begin{array}{c} \psi(i;\zeta)\sim\prod_{r=i_o}^{i-1}\left(\frac{p_r-\zeta}{p_r+\zeta}\right)^{\frac12} \\ \rmpsi(i;\zeta)\sim\prod_{r=i_o}^{i-1}\left(\frac{p_r+\zeta}{p_r-\zeta}\right)^{\frac12} \end{array} \right\} \; {\rm as} \; i\to+\infty,
\end{equation}
where $i_o$ is some constant. 
\end{definition}
Since \eqref{eq:phi1} is invariant under the map $\zeta\to-\zeta$, it follows by the definition of the boundary conditions for the Jost solutions and uniqueness of the boundary value problem \cite{m:90}, that
\begin{equation}
\rvp(i;\zeta)=\vp(i;-\zeta), \;\;\; \rmpsi(i;\zeta)=\psi(i;-\zeta).
\end{equation}
Since the general solution to \eqref{eq:phi1} involves two linearly independent solutions we may write
\begin{equation}\label{eq:ab}
\psi=\az\rvp+\bz\vp, \;\;\; \rmpsi=\mr{\az}\vp+\mr{\bz}\rvp
\end{equation}
where $\az$ and $\bz$ are independent of $i$ and $\mr{\az}(\zeta)=\az(-\zeta)$ and $\mr{\bz}(\zeta)=\bz(-\zeta)$. Moreover one can show that for any two solutions of \eqref{eq:phi1} the Wronskian
\begin{equation}
W(x,y)=(p_i^2-\zeta^2)^{\frac12}(x\ol{y}-\ol{x}y)
\end{equation}
is independent of $i$, and thus by choosing the relevant values at the boundary we have $W(\rvp,\vp)=W(\psi,\rmpsi)=2\zeta$, implying
\begin{equation}
\az\mr{\az}-\bz\mr{\bz}=1.
\end{equation}

\subsection{Analyticity and Asymptotics} $\\$

We now consider the analyticity and asymptotic properties of the Jost solutions and the spectral functions $\az$ and $\bz$. For convenience we make the following definition.

\begin{definition}
The functions $\lam$, $\rlam$ and $\up$, $\rup$ are defined by
\begin{subequations}
\begin{align}
\vp=\lam\prod_{r=i_o}^{i-1}\left(\frac{p_r+\zeta}{p_r-\zeta}\right)^{\frac12},& \;\;\; \rvp=\rlam\prod_{r=i_o}^{i-1}\left(\frac{p_r-\zeta}{p_r+\zeta}\right)^{\frac12} \\
\psi=\up\prod_{r=i_o}^{i-1}\left(\frac{p_r-\zeta}{p_r+\zeta}\right)^{\frac12},& \;\;\; \rmpsi=\rup\prod_{r=i_o}^{i-1}\left(\frac{p_r+\zeta}{p_r-\zeta}\right)^{\frac12}.
\end{align}
\end{subequations}
\end{definition}

For the purposes of this section it is also convenient to scale the potential $\pot$ by introducing a multiplicative scaling factor $\lambda$. That is, we rewrite the potential as
\[
\pot\to\lambda\pot,
\]
where $\lambda$ is a new complex parameter. This parameter is introduced solely for the analysis presented in this section, and thus will not actually appear in the solution $u$. Its benefit however is that it acts as a perturbation (from $\lambda=1$) of the initial-value space treated in the direct scattering problem. This perturbation does not affect the boundary conditions placed on $u$, and is introduced exclusively to analyse under what circumstances a potential $\pot$ will be {\it admissible} (see Definition \ref{def:admis}). The Jost solutions and the spectral functions will (for the present) depend on $\lambda$, and we make this explicit by writing $\lam(i;\zeta)\to\lam(i;\zeta,\lambda)$, etc. 
\par
The linear equation \eqref{eq:phi1} is a generalisation of that treated in \cite{bj:10}, where the direct scattering was carried out along a line rather than a staircase. This changes the parameters $\pp_i$ which are present along $\Gamma$ but does not alter the nature of the scattering problem. Except for the adjustments required to allow for non-constant complex parameter values and the inclusion of the parameter $\lambda$, the proofs of Lemma \ref{lem:jostsum} and Proposition \ref{pr:jostanal} follow those given in \cite{bj:10}. 

\begin{lemma}\label{lem:jostsum}
The functions $\lam$ and $\up$ satisfy the following summation equations\footnote{$F(i,l;\zeta)$ and $G(i,l;\zeta)$ have removable singularities at $\zeta=0$. To make this clear one may rewrite them in a different form, for example if $\Gamma$ spans two lattice directions $n_1$ and $n_2$ we have
\[
F(i,l;\zeta)=\frac1{p_1+\zeta}\sum_{r=0}^{t_1-1}\left(\frac{p_1-\zeta}{p_1+\zeta}\right)^r+\frac1{p_2+\zeta}\sum_{s=0}^{t_2-1}\left(\frac{p_2-\zeta}{p_2+\zeta}\right)^s
\]
for some $t_1$ and $t_2$ satisfying $t_1+t_2=i-l$.}:
\begin{equation}\label{eq:jostsum}
\lam(i;\zeta,\lambda)=1+\lambda\sum_{l=-\infty}^{i-1}F(i,l;\zeta)\pot_l\lam(l;\zeta,\lambda), \;\;\;\;\; \up(i;\zeta,\lambda)=1+\lambda\sum_{l=i+1}^{+\infty}G(i,l;\zeta)\pot_l\up(l;\zeta,\lambda),
\end{equation}
where
\[
F(i,l;\zeta)=\frac1{2\zeta}\left[1-\prod_{r=l}^{i-1}\left(\frac{p_r-\zeta}{p_r+\zeta}\right)\right], \;\;\;\;\; G(i,l;\zeta)=\frac1{2\zeta}\left[1-\prod_{r=i}^{l-1}\left(\frac{p_r-\zeta}{p_r+\zeta}\right)\right].
\]
These summation equations have the Neumann series solutions
\begin{equation}\label{eq:jostseries}
\lam(i;\zeta,\lambda)=\sum_{k=0}^{+\infty}\lambda^kH_k(i;\zeta), \;\;\;\;\; \up(i;\zeta,\lambda)=\sum_{k=0}^{+\infty}\lambda^kJ_k(i;\zeta)
\end{equation}
where
\[
H_0=1, \;\;\;\;\; H_{k+1}(i;\zeta)=\sum_{l=-\infty}^{i-1}F(i,l;\zeta)\pot_lH_k(l;\zeta), \;\;\;\;\; J_0=1, \;\;\;\;\; J_{k+1}(i;\zeta)=\sum_{l=i+1}^{+\infty}G(i,l;\zeta)\pot_lJ_k(l;\zeta).
\]
\end{lemma}

\begin{proposition}\label{pr:jostanal}
Let ${\rm I}$ denote the collection of lattice directions in which $\Gamma $ exists, and assume that 
\begin{equation}\label{eq:wsum}
\sum_{i=-\infty}^{+\infty}|\pot_i|\cK_o^{|i|}(1+|i|)<\infty
\end{equation}
where
\[
\cK_o=\cK_o(a,b)>\max_{r\in{\rm I}}\left\{\left|\frac{p_r+a}{p_r-a}\right|, \left|\frac{p_r-a}{p_r+a}\right|, \left|\frac{p_r+b}{p_r-b}\right|, \left|\frac{p_r-b}{p_r+b}\right|\right\}\geq1.
\]
Define the two open regions $\cP$ and $\cP^*$ by
\[
\cP:=\bigcup_{r\in{\rm I}}\left\{\zeta\;:\;\left|\frac{p_r-\zeta}{p_r+\zeta}\right|>\cK_o\right\}, \;\;\;\;\; \cP^*:=\bigcup_{r\in{\rm I}}\left\{\zeta\;:\;\left|\frac{p_r+\zeta}{p_r-\zeta}\right|>\cK_o\right\}\footnote{Since $\cK_o>1$ both $\cP$ and $\cP^*$ are bounded open regions with empty intersection. As an example, in the case of $\Gamma$ being a line in the $n$-direction we have $\cP=\{\zeta=x+iy:(x+\lambda p_{re})^2+(y+\lambda p_{im})^2<(\lambda^2-1)|p|^2\}$, $\cP^*=\{\zeta=x+iy:(x-\lambda p_{re})^2+(y-\lambda p_{im})^2<(\lambda^2-1)|p|^2\}$, where $p=p_{re}+ip_{im}$ and $\lambda=\frac{\cK_o^2+1}{\cK_o^2-1}$.},
\]
then for $\zeta\notin\cP$ and $\lambda\in D=\Bigl\{\lambda:|\lambda-1|<d\Bigr\}$, $d>0$, the series solutions \eqref{eq:jostseries} converge absolutely. For each $i$ the convergence is uniform in $\zeta$ and $\lambda$, and thus $\lam$ and $\up$ are analytic in $\zeta$ for $\zeta\notin\{\cP\cup\partial\cP\}$, analytic in $\lambda$ for $\lambda\in D$ and continuous in $\zeta$ for $\zeta\notin\cP$. Equivalent results hold for $\rlam$ and $\rup$ for $\zeta\notin\cP^*$. 
\end{proposition}

The summability condition \eqref{eq:wsum} is the only restriction that we impose on the initial profile, i.e. the potential $\pot$. This is a stronger condition that one finds for such partial differential equations as the KdV equation. The ${\rm Q}3_\delta$ equation represents a different class of equations to the KdV, and thus one should not expect that \eqref{eq:wsum} be equivalent to that obtained for the KdV equation \cite{dt:79} or the lattice potential KdV equation \cite{bj:10}. It is shown in Section \ref{sec:degen} however that by the degeneration scheme given in \cite{nah:09} one can obtain solutions to all lower ABS equations, including the lattice potential KdV equation (H1) and the lattice KdV equation. This degeneration process alters the summability condition \eqref{eq:wsum}, and for these equations this condition reduces to
\[
\sum_{i=-\infty}^{+\infty}|\pot_i|(1+|i|)<\infty,
\]
which is equivalent to that imposed for the KdV partial differential equation \cite{dt:79} and the lattice potential KdV equation \cite{bj:10}. For these reasons the condition \eqref{eq:wsum} is not restrictive in the sense that it only allows the reproduction of the soliton solutions, but rather allows for a wider class of solutions with background radiation to be found. 

\begin{corollary}
\begin{equation}\label{eq:absum}
\az(\zeta,\lambda)=1+\frac\lambda{2\zeta}\sum_{l=-\infty}^{+\infty}\pot_l\up(l;\zeta,\lambda), \;\;\;\;\; \bz(\zeta,\lambda)=-\frac\lambda{2\zeta}\sum_{l=-\infty}^{+\infty}\left[\prod_{r=i_o}^{l-1}\left(\frac{p_r-\zeta}{p_r+\zeta}\right)\right]\pot_l\up(l;\zeta,\lambda).
\end{equation}
$\az$ and $\az_\lambda$ are continuous in $\zeta$ for $\zeta\notin\cP$, except possibly at $\zeta=0$, and analytic in $\zeta$ in the interior of this region. $\bz$ and $\bz_\lambda$ are continuous in $\zeta$ for $\zeta\notin\{\cP\cup\cP^*\}$, except possibly at $\zeta=0$, and analytic in $\zeta$ in the interior of this region. Both are analytic in $\lambda$ for $\lambda\in D$. 
\begin{proof}
\eqref{eq:absum} follow directly from comparing \eqref{eq:jostsum} with
\[
\up\sim\az+\bz\prod_{r=i_o}^{i-1}\left(\frac{p_r+\zeta}{p_r-\zeta}\right) \; {\rm as} \; i\to-\infty.
\]
Taking the Wronskian of \eqref{eq:ab} with respect to $\vp$ and $\rvp$ gives
\begin{subequations}
\begin{align}\label{eq:awronskian}
\az&=\frac1{2\zeta}W(\psi,\vp)=\frac1{2\zeta}\Bigl[(p_i+\zeta)\up(i;\zeta,\lambda)\lam(i+1;\zeta,\lambda)-(p_i-\zeta)\up(i+1;\zeta,\lambda)\lam(i;\zeta,\lambda)\Bigr] \\
\label{eq:bwronskian}
\bz&=\frac1{2\zeta}W(\rvp,\psi)=\left(\frac{p_i-\zeta}{2\zeta}\right)\prod_{r=i_o}^{i-1}\left(\frac{p_r-\zeta}{p_r+\zeta}\right)\Bigl[\rlam(i;\zeta,\lambda)\up(i+1;\zeta,\lambda)-\rlam(i+1;\zeta,\lambda)\up(i;\zeta,\lambda)\Bigr].
\end{align}
\end{subequations}
The remaining statements follow from these relations and Proposition \ref{pr:jostanal}.
\end{proof}
\end{corollary}

\begin{lemma}
For $\lambda\in D$, 
\begin{align}\label{eq:jostinf}
&\lam=1+O\left(\frac1{\zeta}\right) \; {\rm as} \; |\zeta|\to\infty, \;\;\;\;\; \up=1+O\left(\frac1{\zeta}\right) \; {\rm as} \; |\zeta|\to\infty, \\
\label{eq:abinf}
&\az=1+O\left(\frac1{\zeta}\right) \; {\rm as} \; |\zeta|\to\infty, \;\;\;\;\; \bz=O\left(\frac1{\zeta}\right) \; {\rm as} \; |\zeta|\to\infty.
\end{align}
\begin{proof}
From the series solutions \eqref{eq:jostseries} it is straightforward to verify that $H_k(i;\zeta)=O\left(\frac1{\zeta^k}\right)$ and $J_k(i;\zeta)=O\left(\frac1{\zeta^k}\right)$ as $|\zeta|\to\infty$, which proves \eqref{eq:jostinf}. To prove the results about $\az$ and $\bz$ we use \eqref{eq:jostinf} in the expressions \eqref{eq:absum}. As $|\zeta|\to\infty$ we have
\begin{align*}
&\az=1+\frac\lambda{2\zeta}\sum_{l=-\infty}^{+\infty}\pot_l\left(1+O\left(\frac1{\zeta}\right)\right)=1+O\left(\frac1{\zeta}\right) \\
&\bz=-\frac{\lambda}{2\zeta}\left(\sum_{l=-\infty}^{+\infty}\left[\prod_{r=i_o}^{l-1}\left(\frac{p_r+\zeta}{p_r-\zeta}\right)\right]\pot_l\right)\left(1+O\left(\frac1{\zeta}\right)\right)=O\left(\frac1{\zeta}\right),
\end{align*}
since for $\zeta\notin\{\cP\cup\cP^*\}$,
\[
\left|\sum_{l=-\infty}^{+\infty}\left[\prod_{r=i_o}^{l-1}\left(\frac{p_r+\zeta}{p_r-\zeta}\right)\right]\pot_l\right|\leq\sum_{l=-\infty}^{+\infty}|\pot_l|\cK_o^{|l-i_o|}<\infty.
\]
\end{proof}
\end{lemma}

\begin{theorem}\label{thm:a}
Assume that $\az$ does not vanish on $\partial\cP$. Then either
\begin{enumerate}
\item[-] $\az(\zeta,1)$ has a finite number of simple zeroes in $\mathbb{C}\setminus\{\cP\cup\partial\cP\}$, or
\item[-] There exists a punctured open disc $D^*$ around $\lambda=1$ such that for $\lambda\in D^*$, $\az(\zeta,\lambda)$ has a finite number of simple zeroes in $\mathbb{C}\setminus\{\cP\cup\partial\cP\}$.
\end{enumerate}
\begin{proof}
$\az$ is analytic in this region, and since $\az\sim1+O\left(\frac1\zeta\right)$ as $|\zeta|\to\infty$, one can show that no limiting sequence of zeroes exists and thus $\az$ must have a finite number of isolated zeroes in this region. For each such zero $\zeta_j$ there exists a locally convergent taylor series
\[
\az(\zeta,\lambda)=\sum_{l=1}^{+\infty}a_l(\lambda)(\zeta-\zeta_j(\lambda))^l
\]
where each $\zeta_j$ and each $a_l$ are analytic in $\lambda$. If $a_1(1)\neq0$ then by continuity $a_1(\lambda)\neq0$ in an open disc $D_j$ around $\lambda=1$. If $a_1(1)=0$ then since $a_1(\lambda)$ is analytic there again exists such a $D_j$ on which $a_1(\lambda)\neq0$ for $\lambda\neq1$. Letting $D^*$ be the intersection of all the sets $D_j$ proves the result.
\end{proof}
\end{theorem}

Theorem \ref{thm:a} illustrates that the potentials which give rise to $\az(\zeta,1)$ having zeroes of order greater than unity in $\mathbb{C}\setminus\{\cP\cup\partial\cP\}$ are exceptional cases, as the potential can simply be rescaled to avoid this circumstance. We therefore now ignore the parameter $\lambda$ altogether, and omit it from the notation, but make the following definition.

\begin{definition}\label{def:admis}
Given a staircase $\Gamma$, the potential $\pot$ is called {\bf admissible} if it satisfies 
\[
\sum_{i=-\infty}^{+\infty}|\pot_i|\cK_o^{|i|}(1+|i|)<\infty,
\]
and gives rise to a spectral function $\az$ which has a finite number of simple zeroes in $\mathbb{C}\setminus\{\cP\cup\partial\cP\}$. 
\end{definition}

Henceforth we assume that the potential has been scaled appropriately so that it is admissible.

\begin{corollary}
Let $\zeta_j$ be a zero of $\az$ in $\mathbb{C}\setminus\{\cP\cup\partial\cP\}$. If $\zeta_j\notin\cP^*$ then $\psi(i;\zeta_j)=\bz(\zeta_j)\vp(i;\zeta_j)$. If $\zeta_j\in\cP^*$ then $\psi(i;\zeta_j)=c_j\vp(i;\zeta_j)$ for some constant $c_j$. 
\begin{proof}
If $\zeta_j\notin\cP^*$ then the result follows from $\psi=\az\rvp+\bz\vp$. If $\zeta_j\in\cP^*$ then by \eqref{eq:awronskian} we have $W(\psi,\vp)=0$, which implies that these two functions are linearly dependent.
\end{proof}
\end{corollary}

\subsection{Multidimensional Dependence of the Spectral Functions}\label{sec:md} $\\$

We now consider how the Jost functions and the spectral functions $\az$ and $\bz$ vary as functions of the lattice parameters $n_1$, ..., $n_{\rm N}$. By definition the functions $\az$ and $\bz$ are independent of $i$, and by making the natural extension
\[
\prod_{r=i_o}^{i-1}\left(\frac{p_r+\zeta}{p_r-\zeta}\right)\to\prod_{r\in{\rm I}}\left(\frac{p_r+\zeta}{p_r-\zeta}\right)^{n_r},
\]
these functions are independent of every variable through which $i$ cycles. For example if N=3 and we have a (1,1) staircase in the $(n,m)$-plane, then $\az$ and $\bz$ will be independent of $n$ and $m$. These functions may however have dependence on the remaining independent variable $l$. Consider one of the remaining ${\rm N}-{\rm I}$ directions in which $\Gamma$ does not exist, with independent variable $n_s$ and parameter $\pp_s$. From \eqref{eq:phi0} the evolution of $\phi_1$ in this direction is governed by
\begin{equation}\label{eq:ds'}
(p_s^2-\zeta^2)^{\frac12}\accentset{\diamond}{\accentset{\diamond}{\phi}}_1-\left(\frac{P_s(\accentset{\diamond}{\accentset{\diamond}{u}}-u)}{\accentset{\diamond}{\cU}}\right)\accentset{\diamond}{\phi}_1+(p_s^2-\zeta^2)^{\frac12}\phi_1=0,
\end{equation}
where $\diamond$ denotes an iteration in the $n_s$-direction. In accordance with the soliton solution behaviour \cite{nah:09}, we assume that in each of these remaining N-I lattice directions,
\begin{equation}\label{eq:otherdirecbc}
\frac{P_s(\accentset{\diamond}{\accentset{\diamond}{u}}-u)}{\accentset{\diamond}{\cU}}\sim2p_s \; {\rm as} \; n_s\to\pm\infty,
\end{equation}
which implies that as $n_s\to\pm\infty$,
\[
(p_s^2-\zeta^2)^{\frac12}\accentset{\diamond}{\accentset{\diamond}{\phi}}_1-2p_s\accentset{\diamond}{\phi}_1+(p_s^2-\zeta^2)^{\frac12}\phi_1=0.
\]
The boundary conditions \eqref{eq:BCjost} defining the Jost functions however are not dependent on $n_s$, and are thus inconsistent with this equation. 

\begin{definition}
The Jost functions which are consistent with \eqref{eq:ds'} in each of the remaining ${\rm N}-{\rm I}$ lattice directions (indexed by the set ${\rm J}$) are defined by
\begin{align*}
&\vp^{({\rm N})}=\vp\prod_{s\in{\rm J}}\left(\frac{p_s+\zeta}{p_s-\zeta}\right)^{\frac12n_s}, \;\;\; \rvp^{({\rm N})}=\rvp\prod_{s\in{\rm J}}\left(\frac{p_s-\zeta}{p_s+\zeta}\right)^{\frac12n_s}, \\
&\psi^{({\rm N})}=\psi\prod_{s\in{\rm J}}\left(\frac{p_s-\zeta}{p_s+\zeta}\right)^{\frac12n_s}, \;\;\; \rmpsi^{({\rm N})}=\rmpsi\prod_{s\in{\rm J}}\left(\frac{p_s+\zeta}{p_s-\zeta}\right)^{\frac12n_s}.
\end{align*}
\end{definition}

We now determine the dependence of $\az$, $\bz$ and the constants $c_j$ on the remaining variables $n_s$ for all $s\in{\rm J}$.

\begin{proposition}
The function $\az$ is independent of $n_s$ for all $s\in{\rm J}$:
\begin{equation}\label{eq:amulti}
\az(n_s;\zeta)=\az(\zeta).
\end{equation}
The dependence of the function $\bz$ and the constants $c_j$ on $n_s$, $s\in{\rm J}$, is given by
\begin{equation}\label{eq:bmulti}
\bz(n_s;\zeta)=\bz(\zeta)\prod_{s\in{\rm J}}\left(\frac{p_s+\zeta}{p_s-\zeta}\right)^{n_s}, \;\;\;\;\; c_j(n_s)=c_j\prod_{s\in{\rm J}}\left(\frac{p_s+\zeta}{p_s-\zeta}\right)^{n_s}
\end{equation}
\begin{proof}
For any given lattice variable $n_s$, $s\in{\rm J}$, the functions $\vp^{(N)}$ and $\rvp^{(N)}$ are linearly independent solutions of \eqref{eq:ds'}, so we may write
\[
\psi^{({\rm N})}=\alpha\rvp^{({\rm N})}+\beta\vp^{({\rm N})} \; \Rightarrow \; \psi=\alpha\rvp+\beta\vp\prod_{s\in J}\left(\frac{p_s+\zeta}{p_s-\zeta}\right)^{n_s},
\]
where $\alpha$ and $\beta$ are independent of $n_k$. Comparing this with $\psi=\az\rvp+\bz\vp$ shows that $\az=\alpha$ is independent of $n_s$, while the dependence of $\bz$ on $n_s$ is a multiplicative factor of the form $\left(\frac{p_s+\zeta}{p_s-\zeta}\right)^{n_s}$. At $\zeta=\zeta_j$ with $\az(\zeta_j)=0$, we have $\psi=c_j\vp$ and thus $c_j$ has the same dependence on $n_s$ as $\bz$. Carrying out this argument in all ${\rm N}-{\rm I}$ directions gives the required result.
\end{proof}
\end{proposition}

Equations \eqref{eq:amulti} and \eqref{eq:bmulti} are the main results of this subsection, and will be used in the following subsection where we carry out the inverse problem.

\subsection{Inverse Problem} $\\$

We now consider the inverse problem, that is the reconstruction of the solution $u$ to $Q3_\delta$. In order to do this we first derive a singular integral equation for the N-dimensional Jost function $\lam$. Since we are considering this function as one depending not just on $i$, but on {\it all} lattice variables $n_1$ to $n_{\rm N}$ (i.e. $\lam=\lam(n_1,...,n_{\rm N};\zeta)$), we must utilise the multivariable dependence \eqref{eq:bmulti} of $\bz$ and the constants $c_j$. By \eqref{eq:amulti} however, the function $\az$ is independent of all lattice variables. Furthermore since the scattering problem is symmetric in all lattice directions, there is no underlying ``preference" given to any particular lattice variable\footnote{this is not the case for partial differential equations such as the KdV equation, where there is a distinct difference between the dependence on $x$ and $t$}. Moreover since the inverse problem is carried out in the complex plane of the spectral parameter $\zeta$ only, it is convenient to alter the notation of the Jost functions by writing
\[
\lam(i;\zeta)\to\lam(\zeta).
\] 
Here it is understood that $\lam$ still depends on all lattice variables $n_1$, ..., $n_{\rm N}$, however for notational reasons we denote only the dependence on the spectral parameter $\zeta$.

\begin{proposition}\label{prop:lamint}
Let $L$ be a closed contour in $\mathbb{C}\setminus\{\cP\cup\partial\cP\}$ defining an interior bounded region $\cR^+$ and an exterior unbounded region $\cR^-$. Assume that $\cP^*\subseteq\cR^+$, $0\in\cR^-$, that all the zeroes of $\az$ in $\mathbb{C}\setminus\{\cP\cup\partial\cP\}$ are contained in $\cR^+$, and that all four points $\zeta=\pm a, \pm b$ lie in $\cR^-$. Then given an admissible potential $\pot_i$ on $\Gamma$, for $\zeta\in\{\cR^-\setminus\cP\}$ the ${\rm N}$-dimensional Jost solution $\lam(\zeta)$ is the unique solution to the singular integral equation
\begin{equation}\label{eq:lamint}
\lam(\zeta)=1-\sum_{j=1}^M\left(\frac{C_j\lam(\zeta_j)}{(\zeta+\zeta_j)}\right)\rho(\zeta_j)+\frac1{2\pi i}\int_{L}\left(\frac{R(\sigma)\lam(\sigma)}{(\sigma+\zeta)}\right)\rho(\sigma)\;d\sigma.
\end{equation}
where the sum is over the $M$ zeroes of $\az$ in $\cR^+$, the constants $C_j$ are given by
\[
C_j:=\frac{c_j}{\az_\zeta(\zeta_j)} \; {\rm if} \; \zeta_j\in\cP^*, \;\;\;\;\; C_j:=\frac{\bz(\zeta_j)}{\az_\zeta(\zeta_j)} \; {\rm if} \; \zeta_j\notin\cP^*,
\]
and the reflection coefficient and plane-wave factors are given respectively by
\[
R(\zeta)=\frac{\bz(\zeta)}{\az(\zeta)}, \;\;\;\;\; \rho(\zeta):=\prod_{r=1}^{{\rm N}}\left(\frac{p_r+\zeta}{p_r-\zeta}\right)^{n_r}.
\]
\begin{proof}
Consider the relation \eqref{eq:ab} for $\lam$ and $\up$:
\[
\frac{\up(\zeta)}{\az(\zeta)}-\rlam(\zeta)=\lam(\zeta)\left(\frac{\bz(\zeta)}{\az(\zeta)}\right)\prod_{r=i_o}^{i-1}\left(\frac{p_r+\zeta}{p_r-\zeta}\right)\prod_{s\in{\rm J}}\left(\frac{p_s+\zeta}{p_s-\zeta}\right)^{n_s}.
\]
By choosing $i_o$ appropriately the plane-wave factor on the right may be rewritten as
\begin{equation}\label{eq:pwf}
\prod_{r=i_o}^{i-1}\left(\frac{p_r+\zeta}{p_r-\zeta}\right)\prod_{s\in{\rm J}}\left(\frac{p_s+\zeta}{p_s-\zeta}\right)^{n_s}=\prod_{r\in{\rm I}}\left(\frac{p_r+\zeta}{p_r-\zeta}\right)^{n_r}\prod_{s\in{\rm J}}\left(\frac{p_s+\zeta}{p_s-\zeta}\right)^{n_s}=\prod_{r=1}^{{\rm N}}\left(\frac{p_r+\zeta}{p_r-\zeta}\right)^{n_r},
\end{equation}
and thus we have $\frac{\up(\zeta)}{\az(\zeta)}-\rlam(\zeta)=\lam(\zeta)R(\zeta)\rho(\zeta)$. The equation
\begin{equation}\label{eq:jump}
\left[\frac{\up(\zeta)}{\az(\zeta)}-1-\sum_{j=1}^M\left(\frac{\up(\zeta_j)}{\az_\zeta(\zeta_j)(\zeta-\zeta_j)}\right)\right]-\left[\rlam(\zeta)-1-\sum_{j=1}^M\left(\frac{\up(\zeta_j)}{\az_\zeta(\zeta_j)(\zeta-\zeta_j)}\right)\right]=\lam(\zeta)R(\zeta)\rho(\zeta),
\end{equation}
then defines a jump condition between two sectionally holomorphic functions along the positively-oriented contour $L$. Given boundary conditions, the question of determining a function $\Phi(\zeta)$ equal to these two functions on $\cR^+$ and $\cR^-$, which also satisfy the jump condition \eqref{eq:jump} on $L$, is a Riemann-Hilbert problem. Since the function defined on $\cR^-$ vanishes as $|\zeta|\to\infty$, the unique solution of this problem (see e.g. Gakhov \cite{g:66} ch.II) is
\[
\Phi(\zeta)=\frac1{2\pi i}\int_{L}\left(\frac{R(\sigma)\lam(\sigma)}{\sigma-\zeta}\right)\rho(\sigma)\;d\sigma.
\]
Using $\up(\zeta_j)=c_j\lam(\zeta_j)\rho(\zeta_j)$ or $\up(\zeta_j)=\bz(\zeta_j)\lam(\zeta_j)\rho(\zeta_j)$ depending on whether or not $\zeta_j\in\cP^*$, the solution $\Phi(\zeta)$ for $\zeta\in\cR^-$ gives \eqref{eq:lamint}. The existence and uniqueness of the solution to \eqref{eq:lamint} follow from the existence and uniqueness properties of the corresponding discrete Gel'fand Levitan integral equation  \cite{bj:10} that one may derive from this equation. In this scenario however we choose to use \eqref{eq:lamint} rather than the Gel'fand-Levitan equation as the solution $u$ of \eqref{eq:Q3} is most naturally expressible in terms of $\lam$.
\end{proof}
\end{proposition}

\begin{remark}
If all of the zeroes of $\az$ do not lie in $\cP^*$ then consider a second positively-oriented contour $L^*$ lying inside $\cR^+$ such that the $M$ zeroes of $\az$ lie between contours $L^*$ and $L$. By the residue theorem we have
\[
\frac1{2\pi i}\left(\int_{L}-\int_{L^*}\right)\left(\frac{R(\sigma)\lam(\sigma)}{(\sigma+\zeta)}\right)\rho(\sigma)\;d\sigma=\sum_{j=1}^M\left(\frac{\bz(\zeta_j)\lam(\zeta_j)}{\az_\zeta(\zeta_j)(\zeta+\zeta_j)}\right)\rho(\zeta_j),
\]
which allows one to rewrite \eqref{eq:lamint} as
\[
\vp(\zeta)+i\rho(\zeta)\int_{L^*}\frac{\vp(\sigma)}{\sigma+\zeta}\;d\lambda(\sigma)=\rho(\zeta),
\]
where the measure $d\lambda(\sigma)$ is given by $2\pi d\lambda(\sigma)=R(\sigma)d\sigma$. By taking the continuum limit in which $\rho(\zeta_k)\to e^{ikx+ik^3t}$, this integral equation becomes the linearisation of the KdV and Painlev\'e II equations, first presented in \cite{fa:81}. This was used as the starting point for the direct linearization method \cite{nqc:83}, and has since seen many generalisations and applications to other nonlinear partial differential and partial difference equations (see e.g. \cite{nqlc:83} and \cite{qncl:84}). 
\end{remark}

We are now in a position to state the main result of this section, that is the reconstruction of the solution $u$ of ${\rm Q}3_\delta$. The solution is expressed in terms of the Jost solution $\lam$ which may be calculated from \eqref{eq:lamint}. 

\begin{theorem}\label{thm:u}
Given an admissible initial condition $\pot_i$ on the staircase $\Gamma$ that is symmetric in its dependence on the parameters $a$ and $b$, the ${\rm N}$-dimensional solution $u$ of ${\rm Q}3_\delta$ is given by
\begin{equation}\label{eq:uQ3}
u=\cA\cF(a,b)\cS_\cA+\cB\cF(a,-b)\cS_\cB+\cC\cF(-a,b)\cS_\cC+\cD\cF(-a,-b)\cS_\cD,
\end{equation}
where $\cF(a,b)$ is defined by \eqref{eq:F} and 
\begin{equation}
\cS_\cA=\cS(a,b), \;\;\; \cS_\cB=\cS(a,-b), \;\;\; \cS_\cC=\cS(-a,b), \;\;\; \cS_\cD=\cS(-a,-b)
\end{equation}
with
\begin{equation}\label{eq:Ssol}
\cS(a,b)=\frac1{a-b}\Bigl((p_k+a)\asf{\lam}(a)\lam(b)-(p_k+b)\lam(a)\asf{\lam}(b)\Bigr),
\end{equation}
where $\frown$ denotes an iteration in any one of the ${\rm N}$ lattice directions, and $n_k$ and $\pp_k$ are the independent variable and associated parameter in this direction. 
\begin{proof}
Consider the two components of any one of the ${\rm N}$ Lax equations of the form \eqref{eq:ds''}:
\begin{subequations}
\begin{align}\label{eq:pf1}
(p_k^2-\zeta^2)^{\frac12}\asf{\phi}_1&=\left(\frac{P_k\asf{u}-(p_k^2-b^2)u}{\cU}\right)\phi_1+\left(\frac{\zeta^2-b^2}{\cU}\right)\phi_2 \\
\label{eq:pf2}
(p_k^2-\zeta^2)^{\frac12}\asf{\phi}_2&=\left(\asf{\cU}-\frac{\delta^2(p_k^2-b^2)}{4P_k\cU(\zeta^2-b^2)}\right)\phi_1+\left(\frac{(p_k^2-b^2)\asf{u}-P_ku}{\cU}\right)\phi_2.
\end{align}
\end{subequations}
The equation governing $\phi_1$ in the $n_k$-direction is
\begin{equation}\label{eq:pf3}
(p_k^2-\zeta^2)^{\frac12}\asf{\asf{\phi}}_1-\left(\frac{P_k(\asf{\asf{u}}-u)}{\asf{\cU}}\right)\asf{\phi}_1+(p_k^2-\zeta^2)^{\frac12}\phi_1=0.
\end{equation}
Now consider the points $\zeta=\pm b$. At these points the equation governing $\phi_2$ in the $n_k$-direction is
\[
(p_k^2-a^2)^{\frac12}\asf{\asf{\phi}}_2-\left(\frac{P_k(\asf{\asf{u}}-u)}{\asf{\cU}}\right)\asf{\phi}_2+(p_k^2-a^2)^{\frac12}\phi_2=0,
\]
thus at $\zeta=b$ we may write
\begin{align*}
\Bigl[(\zeta^2-b^2)\vp^{(N)}_2(\zeta)\Bigr]_{\zeta=b}\;&=c_1\vp^{(N)}(a)+c_2\vp^{(N)}(-a) \\
\Rightarrow \Bigl[(\zeta^2-b^2)\vp_2(\zeta)\Bigr]_{\zeta=b}\;&=c_1\vp(a)\prod_{s\in J}\left(\frac{(p_s+a)(p_s-b)}{(p_s-a)(p_s+b)}\right)^{\frac12n_s}+c_2\vp(-a)\prod_{s\in J}\left(\frac{(p_s-a)(p_s-b)}{(p_s+a)(p_s+b)}\right)^{\frac12n_s}
\end{align*}
for some functions $c_1$ and $c_2$ independent of $n_k$. Since the same argument may be carried out in all lattice directions it follows that $c_1$ and $c_2$ are constant. Using \eqref{eq:pwf}, and the fact that $\cF(a,b)=\rho(a)^{\frac12}\rho(b)^{\frac12}$, equation \eqref{eq:pf1} at $\zeta=b$ then becomes
\begin{equation}\label{eq:pf4}
(p_k+b)\left(\frac{\asf{\lam}(b)}{\lam(b)}\right)\cU=P_k\asf{u}-(p_k^2-b^2)u+c_1\cF(a,-b)\left(\frac{\lam(a)}{\lam(b)}\right)+c_2\cF(-a,-b)\left(\frac{\lam(-a)}{\lam(b)}\right).
\end{equation}
Writing
\[
\cU=\cA\cF(a,b)\cS_\cA^*+\cB\cF(a,-b)\cS_\cB^*+\cC\cF(-a,b)\cS_\cC^*+\cD\cF(-a,-b)\cS_\cD^*
\]
and taking the coefficient of $\cF(a,b)$ in \eqref{eq:pf4} one obtains
\[
(p_k+b)\left(\frac{\asf{\lam}(b)}{\lam(b)}\right)=\left(\frac{(p_k+a)(p_k+b)\asf{\cS}_\cA-(p_k^2-b^2)\cS_\cA}{\cS_\cA^*}\right).
\]
A crucial observation is that one is free to interchange the roles of $a$ and $b$ in the ${\rm N}$ lax matrices \eqref{eq:Nlax}, and thus we also have
\[
(p_k+a)\left(\frac{\asf{\lam}(a)}{\lam(a)}\right)=\left(\frac{(p_k+a)(p_k+b)\asf{\cS}_\cA-(p_k^2-a^2)\cS_\cA}{\cS_\cA^*}\right).
\]
Combining these gives the two expressions
\begin{subequations}\label{eq:pf5}
\begin{align}
&(a^2-b^2)\left(\frac{\cS_\cA}{\cS_\cA^*}\right)=\frac{(p_k+a)\asf{\lam}(a)\lam(b)-(p_k+b)\lam(a)\asf{\lam}(b)}{\lam(a)\lam(b)} \\
&(a^2-b^2)\left(\frac{\asf{\cS}_\cA}{\cS_\cA^*}\right)=\frac{(p_k-b)\asf{\lam}(a)\lam(b)-(p_k-a)\lam(a)\asf{\lam}(b)}{\lam(a)\lam(b)}.
\end{align}
\end{subequations}
One final Wronskian identity results from considering equation \eqref{eq:pf3} for $\lam(a)$ and $\lam(b)$. By eliminating the middle term one obtains
\[
(p_k-b)\asf{\lam}(a)\lam(b)-(p_k-a)\lam(a)\asf{\lam}(b)=(p_k+a)\asf{\asf\lam}(a)\asf{\lam}(b)-(p_k+b)\asf{\lam}(a)\asf{\asf\lam}(b),
\]
and combining this with \eqref{eq:pf5} gives
\[
\cS_\cA=\frac1{a-b}\Bigl((p_k+a)\asf{\lam}(a)\lam(b)-(p_k+b)\lam(a)\asf{\lam}(b)\Bigr).
\]
By considering the points $\zeta=-b$, $\zeta=a$ and $\zeta=-a$ one can obtain expressions for the remaining functions $\cS_\cB$, $\cS_\cC$ and $\cS_\cD$ in a similar fashion.
\end{proof}
\end{theorem}

This concludes the derivation of the IST for ${\rm Q}3_\delta$. The entire procedure may be summarised as follows: Given a staircase $\Gamma$ in an N-dimensional lattice of the form stipulated in Subsection \ref{subsec:IS}, one then specifies an initial profile $\pot_i$ along this staircase. This potential must satisfy the summability condition \eqref{eq:wsum}. By solving equation \eqref{eq:phi1} one can construct the Jost solutions along $\Gamma$, and thereby determine the functions $\az$ and $\bz$ and the constants $c_j$. Finally one can use the singular integral equation \eqref{eq:lamint} to determine $\lam$ as a function of all lattice variables $n_1,...,n_{\rm N}$ and obtain the solution $u$ of ${\rm Q}3_\delta$ using Theorem \ref{thm:u}. 

\section{Degenerations}\label{sec:degen}

Theorem \ref{thm:u} allows one to express the solution \eqref{eq:uQ3} of ${\rm Q}3_\delta$ in terms of $\lam(\pm a)$ and $\lam(\pm b)$. We now consider the problem of degenerating this solution to the remaining equations in the lists \eqref{eq:Hlist} and \eqref{eq:Qlist}. To do so we follow the degenerations given in \cite{nah:09} which are limits on the parameters $a$ and $b$ and the dependent variable $u$, in which a small parameter $\epsilon$ is introduced, and all degenerations are obtain in the limit $\epsilon\to0$. Further choices must be made for the degenerations of the constants appearing in the solution in order that the resulting solutions be of the required form. These choices are given in \cite{nah:09} and are repeated here for convenience.  The most complicated is the ${\rm Q}3_\delta\to{\rm Q}2$ degeneration, in which we make the substitution $b=a(1-2\epsilon)$. The summability condition \eqref{eq:wsum} on the potential $\pot$ guarantees the existence and differentiability of $\lam$ at $\zeta=\pm a$ and $\zeta\pm b$, and after taking the required limit on $b$ we have
\[
\cK_o(a,b)\to\cK_1(a)>\max_{r\in{\rm I}}\left\{\left|\frac{p_r+a}{p_r-a}\right|, \left|\frac{p_r-a}{p_r+a}\right|\right\},
\]
and thus the solution to ${\rm Q}2$ will satisfy the summability condition
\begin{equation}\label{eq:wsum'}
\sum_{i=-\infty}^{+\infty}|\pot_i|\cK_1^{|i|}(1+|i|)<\infty.
\end{equation}
This is also the case for the solutions of ${\rm Q}1_\delta$ and H3. The solutions to H2 and H1, which are obtained by degenerating the Q2 and ${\rm Q}1_\delta$ solution respectively, require the parameter $a$ to approach infinity. By choosing $\cK_1$ such that $\cK_1(a)\to1$ as $|a|\to\infty$ then the solutions obtained will satisfy the summability condition
\begin{equation}\label{eq:wsum''}
\sum_{i=-\infty}^{+\infty}|\pot_i|(1+|i|)<\infty.
\end{equation}
In the following degenerations we expand the Jost function $\lam(\zeta)$ using Taylor series. For example after setting $b=a(1-2\epsilon)$ in the ${\rm Q}3_\delta\to$ Q2 degeneration one writes
\[
\lam(b)=\lam(a)-2a\lam'(a)\epsilon+2a^2\lam''(a)\epsilon^2+... \; .
\]
Note that in the following the notation $\lam'(-a)$ denotes $\lam'(\zeta)=\frac{{\rm d}}{{\rm d}\zeta}\lam(\zeta)$ evaluated at $\zeta=-a$. In addition to this one must make use of the Wronskian identity
\begin{equation}\label{eq:identity}
(p_k+a)\asf{\lam}(a)\lam(-a)-(p_k-a)\lam(a)\asf{\lam}(-a)=2a,
\end{equation}
obtained by combining equation \eqref{eq:pf3} for $\lam(a)$ and $\lam(-a)$, as well as the two further identities obtained by differentiating \eqref{eq:identity} once and twice with respect to $a$. Degenerations of the exponential terms $\cF$ are easily computed and can be seen in \cite{nah:09}. For the degenerations where the parameters $a$ or $b$ approach infinity we make use of \eqref{eq:jostinf} which gives
\[
\lam\left(\frac1\epsilon\right)\sim1+\epsilon\Bigl(\lam_\epsilon\Bigr)+\epsilon^2\Bigl(\lam_{\epsilon^2}\Bigr)+O\left(\epsilon^3\right) \; {\rm as} \; \epsilon\to0,
\]
where functions $\lam_\epsilon$ and $\lam_{\epsilon^2}$ may be calculated by taking the limit $|\zeta|\to\infty$ of $\lam(\zeta)$. For each degeneration we state the limits that one must take on the various terms and give the final degenerated solution. The various new functions obtained involve shifts in an arbitrary $n_k$-direction, denoted again by $\frown$, with associated parameter $\pp_k$.

\subsection{Q2} $\\$

As a first case we consider the degeneration to Q2 which lies just below ${\rm Q}3_\delta$ in the ABS hierarchy. The degeneration is
\[
b=a(1-2\epsilon), \;\;\;\;\; u\to\frac{\delta}{4a^2}\left(\frac1\epsilon+1+(1+2u)\epsilon\right),
\]
and the choices that we make for the constants in order that the solution degenerates in this manner are
\begin{align*}
&\cA\to\frac{\delta}{4a^2}\Bigl(\cA\epsilon\Bigr), \;\;\;\;\; \cB\to\frac{\delta}{8a^2}\left(\frac1\epsilon+1-\xi_o+\left(\frac{3+\xi_o^2}2+2\cA\cD\right)\epsilon\right), \\
&\cC\to\frac{\delta}{8a^2}\left(\frac1\epsilon+1+\xi_o+\left(\frac{3+\xi_o^2}2+2\cA\cD\right)\epsilon\right), \;\;\;\;\; \cD\to\frac{\delta}{4a^2}\Bigl(\cD\epsilon\Bigr).
\end{align*}
Note that the four constants $\cA$, $\cB$, $\cC$ and $\cD$, constrained by \eqref{eq:abcd} have been replaced by the three new constants $\cA$, $\cD$ and $\xi_o$ with no constraint. After this degeneration one finds the new inverse scattering solution of Q2 to be
\begin{align}\label{eq:uQ2}
u=&\frac14\Bigl((\xi+\xi_o)^2+1\Bigr)+a\Bigl(\xi+\xi_o\Bigr)S(-a,a)+a^2\cZ(a)+\cA\cD \\
&+\frac12\cA\rho(a)\Bigl(1-2aS(a,a)\Bigr)+\frac12\cD\rho(-a)\Bigl(1+2aS(-a,-a)\Bigr)
\end{align}
where
\begin{align*}
S(a,b)&=\frac1{2b}\Bigl(1+(p_k+a)\asf{\lam}(a)\lam'(b)-(p_k+b)\lam(a)\asf{\lam}'(b)-\lam(a)\asf{\lam}(b)\Bigr) \\
\cZ(a)&=\frac1{2a}\Bigl(\bigl[(p_k+a)\asf{\lam}'(a)+\asf{\lam}(a)\bigr]\lam'(-a)-\bigl[(p_k-a)\asf{\lam}'(-a)+\asf{\lam}(-a)\bigr]\lam'(a)\Bigr), \\
\xi&=\xi(a)=2a\sum_{r=1}^N\left(\frac{p_r}{a^2-p_r^2}\right)n_r.
\end{align*}

\subsection{Q1$_\delta$} $\\$

The degeneration from Q2 to ${\rm Q}1_\delta$ is given by
\[
u\to\frac{\delta^2}{4\epsilon^2}+\frac1\epsilon u
\]
and the required degenerations of the constants $\cA$, $\cD$ and $\xi_o$ are
\[
\cA\to\frac{2\cA}\epsilon, \;\;\;\;\; \cD\to\frac{2\cD}\epsilon, \;\;\;\;\; \xi_o\to\xi_o+\frac{2\cB}\epsilon.
\]
One then finds the solution of ${\rm Q}1_\delta$ to be
\begin{equation}\label{eq:uQ1}
u=\cB\Bigl(\xi+\xi_o+2aS(-a,a)\Bigr)+\cA\rho(a)\Bigl(1-2aS(a,a)\Bigr)+\cD\rho(-a)\Bigl(1+2aS(-a,-a)\Bigr),
\end{equation}
where the constants $\cA$, $\cB$, $\cD$ and $\xi_o$ are constrained by 
\[
\cB^2+4\cA\cD=\frac{\delta^2}4.
\]

\subsection{H3$_\delta$} $\\$

To obtain the ${\rm H}3_\delta$ solution we degenerate from the ${\rm Q}3_\delta$ solution \eqref{eq:uQ3} by setting
\[
b=\frac1{\epsilon^2}, \;\;\;\;\; u\to\epsilon^3\left(\frac{\sqrt{\delta}}2u\right),
\]
and
\[
\cA\to\epsilon^3\left(\frac{\sqrt\delta}2\cA\right), \;\;\;\;\; \cB\to\epsilon^3\left(\frac{\sqrt\delta}2\cB\right), \;\;\;\;\; \cC\to\epsilon^3\left(\frac{\sqrt\delta}2\cC\right), \;\;\;\;\; \cD\to\epsilon^3\left(\frac{\sqrt\delta}2\cD\right).
\]
The solution which emerges is
\begin{equation}\label{eq:uH3}
u=\Bigl(\cA+\sigma\cB\Bigr)\vartheta\lam(a)+\Bigl(\sigma\cC+\cD\Bigr)\vartheta^{-1}\lam(-a)
\end{equation}
where
\[
\vartheta=\vartheta(a)=\prod_{r=1}^N\left(\frac{P_r}{a-p_r}\right)^{n_r}, \;\;\;\;\; \sigma=(-1)^{n_1+...+n_N},
\]
and the constants $\cA$, $\cB$, $\cC$ and $\cD$ are constrained by
\[
\cA\cD-\cB\cC=\frac{-\delta}{4a}.
\]

\subsection{H2} $\\$

In order to obtain the H2 solution we choose to degenerate from the Q2 solution \eqref{eq:uQ2} by setting
\[
a=\frac1\epsilon, \;\;\;\;\; u\to\frac14+\epsilon^2u.
\]
The degenerations of the constants $\cA$, $\cD$ and $\xi_o$ are
\[
\cA\to\cA\Bigl(\epsilon+2\varrho_1\epsilon^2\Bigr), \;\;\;\;\; \cD\to\cA\Bigl(-\epsilon+2\varrho_1\epsilon^2\Bigr), \;\;\;\;\; \xi_o\to2\epsilon\varrho_o,
\]
where $\cA$, $\varrho_o$ and $\varrho_1$ are unconstrained constants. The solution of H2 is then
\begin{equation}\label{eq:uH2}
u=\Bigl(\varrho+\varrho_o\Bigr)^2+2\Bigl(\varrho+\varrho_o\Bigr)\lam_\epsilon+2\lam_{\epsilon^2}-\cA^2+\sigma2\cA\Bigl(\varrho+\varrho_1+\lam_\epsilon\Bigr),
\end{equation}
where
\[
\varrho=\sum_{r=1}^Np_rn_r.
\]

\subsection{H1} $\\$

Finally we derive the inverse scattering solution to H1 by degenerating from the ${\rm Q}1_\delta$ solution \eqref{eq:uQ1}. We set
\[
a=\frac1\epsilon, \;\;\;\;\; u\to\epsilon\delta u,
\]
and require that the constants degenerate to
\[
\cB\to\frac\delta2\cB, \;\;\;\;\; \cA\to\frac\delta4\cA\Bigl(1+2\varrho_1\epsilon\Bigr), \;\;\;\;\; \cD\to\frac\delta4\cA\Bigl(-1+2\varrho_1\epsilon\Bigr), \;\;\;\;\; \xi_o\to2\epsilon\varrho_o,
\]
where
\[
\cA^2-\cB^2=-1.
\]
The solution of H1 which emerges is
\begin{equation}\label{eq:uH1}
u=\cB\Bigl(\varrho+\varrho_o+\lam_\epsilon\Bigr)+\sigma\cA\Bigl(\varrho+\varrho_1+\lam_\epsilon\Bigr).
\end{equation}
This solution agrees with that given in \cite{bj:10} however is formulated in terms of the expansion of $\lam(\zeta)$ for large $\zeta$ rather than using the Gel'fand-Levitan approach. 

\section{Reflectionless Potentials}\label{sec:arb}

Consider the case of an arbitrary reflectionless potential, i.e. one for which $R(\zeta)\equiv0$, where $\az$ vanishes at exactly one point $\zeta=k$ in $\cP^*$. The integral equation for $\lam(\zeta)$ becomes
\[
\lam(\zeta)=1-\left(\frac{2k\rho^o\lam(k)}{\zeta+k}\right)\rho(k),
\]
where we have written $C_1=2k\rho^o$. The solution is
\[
\lam(\zeta)=\frac{1+\az(\zeta)\rho^o\rho(k)}{1+\rho^o\rho(k)}
\]
where $\az(\zeta)=\frac{\zeta-k}{\zeta+k}$, and thus by \eqref{eq:Ssol} we have
\[
\cS_\cA=\cS(a,b)=\dfrac{1+\frac{(a-k)(b-k)}{(a+k)(b+k)}\rho^o\rho(k)}{1+\rho^o\rho(k)}.
\]
For the reflectionless case where $\az$ vanishes at $M$ points $k_1, ..., k_M$ in $\cP^*$, one can show that the solution of \eqref{eq:lamint} is
\[
\lam(\zeta)=1-{\bf c}^T\Bigl({\rm I}+\cM\Bigr)^{-1}\Bigl(\zeta\,{\rm I}+K\Bigr)^{-1}{\bf r},
\]
where ${\bf c}^T=\Bigl(C_1, C_2, ..., C_M\Bigr)$, ${\bf r}=\Bigl(\rho(k_1), ..., \rho(k_M)\Bigr)^T$, $K_{ij}=k_i\delta_{ij}$ and the Cauchy matrix $\cM$ is given by
\[
\cM_{ij}=\frac{\rho(k_i)C_j}{k_i+k_j}.
\]
This is in fact the quantity $V(\zeta)$ given in \cite{nah:09}, and from the various identities given in this paper one can show that
\[
\cS_{\cA}=\cS(a,b)={\bf c}^T\Bigl(a\,{\rm I}+K\Bigr)^{-1}\Bigl({\rm I}+\cM\Bigr)^{-1}\Bigl(b\,{\rm I}+K\Bigr)^{-1}{\bf r}.
\]
This is exactly the N-soliton solution obtained by the Cauchy matrix approach in \cite{nah:09}. Since the degenerations follow those from this paper, any reflectionless potential will give rise to pure soliton solutions to all equations in the degeneration scheme, that is all ABS equations below Q4.

\section{Other Lattice Equations}\label{sec:others}

In this section we give a short discussion on how the machinery presented in the previous sections allows one to obtain inverse scattering solutions to some previously known lattice equations, including some which are not multidimensionally consistent. We begin by deriving a closed-form lattice equation for the quantity $\lam(\zeta)$. Consider the evolution equation \eqref{eq:phi0} for $\lam(\zeta)$ taken along opposing sides of a lattice element in the $n$- and $m$-directions:
\begin{subequations}
\begin{align}\label{eq:other1}
(q+\zeta)\frac{\wh{\wt\lam}(\zeta)}{\wt{\lam}(\zeta)}+(p-\zeta)\frac{\lam(\zeta)}{\wt{\lam}(\zeta)}&=\frac{Q\uth-(q^2-p^2)\ut-Pu}{\cUt} \\
\label{eq:other2}
(p+\zeta)\frac{\wh{\wt\lam}(\zeta)}{\wh{\lam}(\zeta)}+(q-\zeta)\frac{\lam(\zeta)}{\wh{\lam}(\zeta)}&=\frac{P\uth-(p^2-q^2)\uh-Qu}{\cUh}
\end{align}
\end{subequations}
One can then show that if $u$ solves ${\rm Q}3_\delta$ then the quantities on the right-hand side of these equation are in fact equal\footnote{this is the (1,2) element of the consistency equation for the Lax pairs in the $n$- and $m$-directions}, and thus equating these shows that $\lam(\zeta)$ satisfies
\begin{equation}\label{eq:lameq}
(p+\zeta)\dfrac{\wh{\wt\lam}(\zeta)}{\wh{\lam}(\zeta)}-(p-\zeta)\dfrac{\lam(\zeta)}{\wt{\lam}(\zeta)}=(q+\zeta)\dfrac{\wh{\wt\lam}(\zeta)}{\wt{\lam}(\zeta)}-(q-\zeta)\dfrac{\lam(\zeta)}{\wh{\lam}(\zeta)}.
\end{equation}
Solutions to this equation exist whenever $\zeta\notin\cP$, provided that the potential $\pot$ satisfies \eqref{eq:wsum}. Let $\alpha_p:=(p^2-\zeta^2)^{\frac12}$, $\alpha_q:=(p^2-\zeta^2)^{\frac12}$, then in terms of the Jost function $\vp$ this becomes
\begin{equation}\label{eq:phieq}
\alpha_p\Bigl(\vp(\zeta)\wh{\vp}(\zeta)-\wt{\vp}(\zeta)\wh{\wt\vp}(\zeta)\Bigr)=\alpha_q\Bigl(\vp(\zeta)\wt{\vp}(\zeta)-\wh{\vp}(\zeta)\wh{\wt\vp}(\zeta)\Bigr),
\end{equation}
that is $\vp(\zeta)$ satisfies the lattice potential modified KdV equation \cite{nqc:83} \cite{qncl:84} with parameters $\alpha_p$ and $\alpha_q$. By setting $W(\zeta):=\dfrac{\wh{\vp}(\zeta)}{\wt{\vp}(\zeta)}$ we then see that $W$ satisfies
\begin{equation}\label{eq:Weq}
\dfrac{\wh{\wt W}(\zeta)}{W(\zeta)}=\dfrac{\bigl(\alpha_p\wh{W}(\zeta)-\alpha_q\bigr)\bigl(\alpha_p-\alpha_q\wt{W}(\zeta)\bigr)}{\bigl(\alpha_p\wt{W}(\zeta)-\alpha_q\bigr)\bigl(\alpha_p-\alpha_q\wh{W}(\zeta)\bigr)},
\end{equation}
which is the lattice modified KdV equation \cite{nc:95}. Equations \eqref{eq:phieq} and \eqref{eq:Weq} are invariant under the transformation $\zeta\to-\zeta$ and thus we have inverse scattering solutions to these equations for all $\zeta$, independent of the choices of the parameters $a$ and $b$. By letting these parameters approach infinity the solutions obtained will give rise to potentials satisfying \eqref{eq:wsum''}. Of course equivalent equations to \eqref{eq:lameq}, \eqref{eq:phieq} and \eqref{eq:Weq} hold in any pair of lattice directions, and each of these equations is multidimensionally consistent. If we now break the covariance between the $n$- and $m$- lattice directions in \eqref{eq:lameq} by setting $\zeta=p$, we have the following lattice equation for the quantity $\lam(p)$:
\begin{equation}\label{eq:V2}
2p=(p+q)\dfrac{\wh{\lam}(p)}{\wt{\lam}(p)}+(p-q)\dfrac{\lam(p)}{\wh{\wt\lam}(p)}.
\end{equation}
The (weaker) multidimensional consistency properties of this equation are discussed in \cite{nah:09}. By setting $\zeta=p$ in \eqref{eq:other1} and $\zeta=q$ in \eqref{eq:other2}, then equating the right-hand sides we have $\wh{\wt\lam}(p)\wh{\lam}(q)=\wh{\wt\lam}(q)\wt{\lam}(p)$ which shows that it is self-consistent to express $\lam(p)$ in terms of the $\tau$-function:
\[
\wt{\lam}(p)=\dfrac f{\wt{f}}, \;\;\;\;\; \wh{\lam}(q)=\dfrac f{\wh{f}}.
\]
Using this in \eqref{eq:V2} gives the 6-point equation
\[
2pf\wh{f}=(p+q)\underaccent{\wtilde}{\wh f}\wt{f}+(p-q)\underaccent{\wtilde}{f}\wh{\wt f}.
\]
By supplementing this with the similar equation obtained by setting $\zeta=q$ in \eqref{eq:lameq}, one can obtain Hirota's discrete-time Toda equation \cite{h:77}:
\[
(p-q)^2\underaccent{\wtilde}{\underaccent{\what}{f}}\wh{\wt f}-(p+q)^2\underaccent{\wtilde}{\wh f}\underaccent{\what}{\wt f}+4pqf^2=0.
\]
We now derive two final lattice equation of KdV type. From the ${\rm Q}1_\delta$ equation, if one makes the degeneration
\[
a=\epsilon, \;\;\;\;\; u\to\epsilon u,
\]
then the equation becomes
\begin{equation}\label{eq:V5}
\dfrac{\bigl(u-\ut\bigr)\bigl(\uh-\uth\bigr)}{\bigl(u-\uh\bigr)\bigl(\ut-\uth\bigr)}=\frac{q^2}{p^2},
\end{equation}
which is the Schwarzian KdV equation \cite{nc:95}, also known as the cross-ratio equation (or ${\rm Q}1_o$). By choosing the constants such that
\[
\cA\to\frac\cB4, \;\;\;\;\; \cB\to-\frac\cB2, \;\;\;\;\; \cD\to\frac\cB4, \;\;\;\;\; \xi_o\to1-2\epsilon\mu_o
\]
with $\mu_o$ constant, the inverse scattering solution to \eqref{eq:V5} is then given by
\begin{equation}\label{eq:V6}
u=\cB\Bigl(\mu+\mu_o-\cT\Bigr),
\end{equation}
where
\[
\cT=\frac p2\Bigl(\wt{\lam}(0)\lam''(0)-\lam(0)\wt{\lam}''(0)\Bigr)-\lam(0)\wt{\lam}'(0), \;\;\;\;\; \mu=\sum_{r=1}^M\frac{n_r}{p_r}.
\]
This solution exists whenever \eqref{eq:wsum'} holds with $\cK_1(0)\neq0$. Alternatively from the solution \eqref{eq:uH1} of H1, also known as the lattice potential KdV equation \cite{h:77} \cite{nqc:83} \cite{qncl:84}, if one defines $\omega:=\ut-\uh$ then the quantity $\omega$ satisfies the lattice KdV equation \cite{h:77}
\begin{equation}\label{eq:latticekdv}
\omega-\wh{\wt\omega}=(p^2-q^2)\left(\dfrac1{\wh\omega}-\dfrac1{\wt\omega}\right),
\end{equation}
to which we have the inverse scattering solution
\begin{equation}
\omega=\Bigl(\cB-(-1)^{n+m}\cA\Bigr)\Bigl(p-q+\wt{\lam}_\epsilon-\wh{\lam}_\epsilon\Bigr),
\end{equation}
which exists provided \eqref{eq:wsum''} holds. Note that equation \eqref{eq:latticekdv} is not multidimensionally consistent.

\section{Conclusion}
 
In this paper we have developed a discrete inverse scattering transform for a hierarchy of integrable lattice equations. At the top of this hierarchy lies ${\rm Q}3_\delta$, for which the discrete IST was carried out. The forward scattering problem was carried out along a multidimensional staircase within an N-dimensional lattice. Even though this staircase may traverse multiple dimensions, the forward scattering problem along the staircase is one-dimensional. Building on the machinery developed in \cite{bj:10} the multidimensional complex-valued solution (depending on N independent variables and N distinct lattice parameters) was shown to be expressible in terms of the quantity $\lam(\zeta)$, which is obtained by solving a singular integral equation which is a discrete analogue of the singular integral equation obtained in the IST for the continuous KdV equation \cite{ggkm:74}.  It was also shown that the soliton solutions presented in \cite{nah:09} correspond precisely with the solutions arising from reflectionless potentials. Following the degeneration scheme given in \cite{nah:09} the solution to ${\rm Q}3_\delta$ was degenerated to give inverse scattering solutions to all remaining ABS equations (except for Q4, which will be addressed in due course), as well as some other previously known integrable lattice equations, including the lattice KdV equation which is not multidimensionally consistent.
\par
The summability condition \eqref{eq:wsum} placed on the potential is the only requirement needed to carry out the discrete IST, provided that it is scaled so as to be admissible in accordance with Definition \ref{def:admis}. This condition changes throughout the degeneration scheme, and the condition placed on the ${\rm Q}3_\delta$ potential reduces to \eqref{eq:wsum''} for the cases of H2 and H1, which agrees with that obtained in the case of the continuous KdV partial differential equation, and the condition found in \cite{bj:10}. 
 
\section{Acknowledgements}

The author would like to thank Frank Nijhoff most sincerely for the time spent at Leeds University in 2011, and the many helpful discussions. The author is also greatly indebted to the continued support and skilled guidance of Nalini Joshi, and the inspiring talks with James Atkinson. This research was funded by the Australian Research Council grant DP0985615, and the Phillip Hofflin International Travel Scholarship, for which the author is most grateful.

\end{document}